\documentclass[3p,times]{elsarticle}
\journal{BENEVOL 2009 - Proceedings of Short Papers}

\usepackage{hyperref}

\title{A Framework for Agile Development of Component-Based Applications}

\author{Guillaume Waignier, Est\'eban Duguep\'eroux, Anne-Fran\c{c}oise Le Meur, Laurence Duchien}

\ead{\{Guillaume.Waignier,Esteban.Dugueperoux,Anne-Francoise.Le\_Meur,Laurence.Duchien\}@inria.fr}

\address{Universit\'e Lille 1 - LIFL CNRS UMR 8022 - INRIA 59650 Villeneuve d'Ascq, France}

\begin{document}


\begin{abstract}

  Agile development processes and component-based software
  architectures are two software engineering approaches that
  contribute to enable the rapid building and evolution of
  applications. Nevertheless, few approaches have proposed a framework
  to combine agile and component-based development, allowing an
  application to be tested throughout the entire development cycle.
  To address this problematic, we have built CALICO, a model-based
  framework that allows applications to be safely developed in an
  iterative and incremental manner. The CALICO approach relies on the
  synchronization of a model view, which specifies the application
  properties, and a runtime view, which contains the application in
  its execution context. Tests on the application specifications that
  require values only known at runtime, are automatically integrated
  by CALICO into the running application, and the captured needed
  values are reified at execution time to resume the tests and inform
  the architect of potential problems. Any modification at the model
  level that does not introduce new errors is automatically propagated
  to the running system, allowing the safe evolution of the
  application.  In this paper, we illustrate the CALICO development
  process with a concrete example and provide information on the
  current implementation of our framework.

\end{abstract}

\maketitle

\section{Introduction}
\label{sec:intro}

In many application domains, software systems need to perpetually and
rapidly evolve to cope with new user and technology requirements.
Being able to modify existing systems or redesign new systems to
rapidly take in account new functionalities or preferences has led to
the proposition of several software engineering approaches such as the
Agile software development methodology~\cite{agile}. One of the key
principles of Agile software development is to build software through
an incremental and iterative process. Each iteration adds a new
feature and produces a fully working system by going through the whole
the software lifecycle, {\em i.e.,} the analyze, develop and test
phases. Another particularity of Agile development is that the testing
activity is not just confined to the classical test phase but rather
integrated throughout the entire lifecycle, meaning that the software
is continuously tested throughout its development, from its
specifications to the final running system, in order to augment the
overall software system quality.

Another software engineering approach that contributes to facilitating
the rapid development of software systems is the use of
component-based software architectures. In this context, the overall
structure of the application is first described with an architecture
description language (ADL)~\cite{Medvidovic2000}. Such description
highlights the needed components and their assembly, which facilitates
the understanding and analysis of the application's properties, such
as behavioral or quality of service properties. If the specifications
are coherent, the application is eventually instantiated, deployed and
executed to be tested.

Although Agile software development and component-based software
engineering (CBSE) may appear quite different approaches, some
works~\cite{Stojanovic-etal-03,xpagile} have identified that both
approaches could benefit to each other, CBSE bringing for example the
capability of building large software and enhancing reusability, and
Agile development offering more flexible development processes for
shorter time-to-market products. Nevertheless we believe that there is
still a bridge between these two approaches, one reason being the lack
of component frameworks that allow incremental and iterative
development processes, as well as throughout-lifecycle testing.

To address this problematic, we have developed a model-based
framework, named CALICO, that enables architects to design and test
component-based systems in an iterative and uniformed
process~\cite{waignier-MODELS-08, waignier-QoSA-09}.
CALICO allows software architects to specify their architectures as
models, and to analyze them with respect to application and platform
constraints.  Our approach enables the testing of the system throughout
the system lifecycle. More concretely, CALICO analyses architecture
models and creates contracts by composing contractual application
properties, {\em e.g.}, behavioral, dataflow, QoS properties. This
composition allows compatible and incompatible interaction to be
identified, as well as partially compatible interactions, which
require runtime checking~\cite{waignier-QoSA-08}. When runtime
checking is needed, CALICO automatically instruments the application
to reify runtime information to complete the resolution of the
partially compatible interaction contract and thus detects if the
given interaction may lead to an error.  By using this framework in
iterative software design processes, architects get design feedback,
{\em i.e.,} information on identified interaction errors, and can then
modify the models accordingly.  Each modification performed on the
model is propagated to the running system since CALICO ensures the
synchronization between the model and the runtime system, both of which
thus coexist during the whole application development. Furthermore,
the solution offered by CALICO is generic regarding underlying
platforms, allowing component platforms to benefit from all the
analyses integrated into CALICO.

The rest of this paper is organized as follows.
Section~\ref{sec:presentation} gives an overview of the CALICO
iterative and incremental development process.
Section~\ref{sec:example} illustrates with a concrete scenario the
CALICO approach. Finally, Section~\ref{sec:conclusion} provides some
information about the current status of our framework implementation.

\section{CALICO Overview}
\label{sec:presentation}

\begin{figure}[t]
	\centering
		\includegraphics[width=1.0\textwidth]{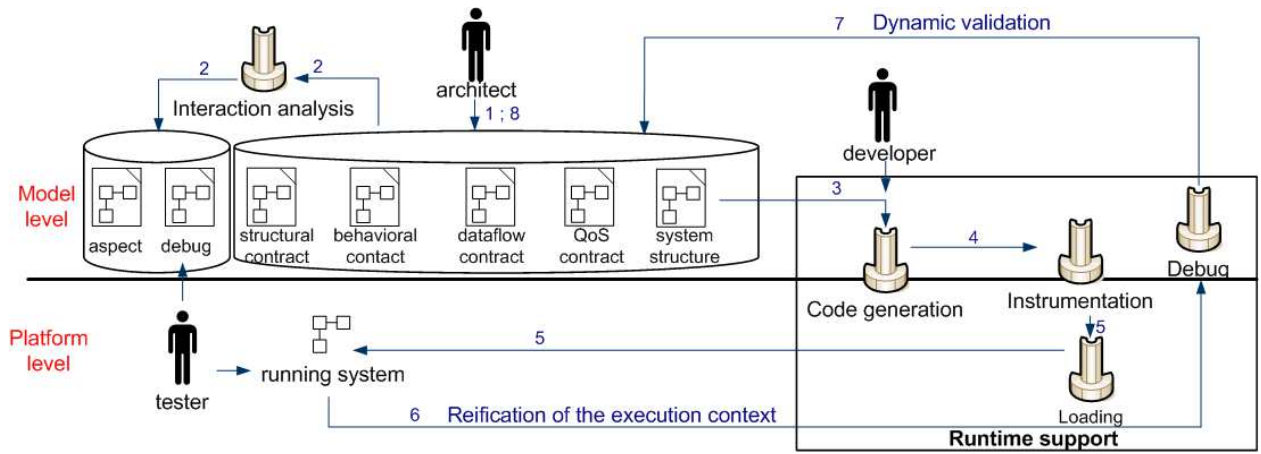}
	\caption{Overview of CALICO iterative development process}
	\label{fig:calico}
\end{figure}

CALICO is composed of two levels: a model level and a platform level
as shown in Figure~\ref{fig:calico}.  The model level is independent
of any component-based or service-oriented platform.  It contains the
CALICO Architecture Description metamodels that enable an architect to
describe the structure and the properties, \textit{i.e.}, structural,
behavioral, dataflow and QoS properties, of an application. It is also
possible to specify some contextual adaptation rules, independently of
any platform, in order to allow the debugging of autonomic systems.  The
platform level holds the running system on a target platform. 

The iterative and incremental development process of CALICO,
illustrated in Figure~\ref{fig:calico}, is as follows :

{\bf (1) Design :} The architect specifies the design of the desired
application using the CALICO metamodels. The {\tt system structure}
metamodel enables architects to describe the structure of their
architecture, independently of any component platform. CALICO provides
also four contract metamodels to allow architects to specify
structural, behavioral, dataflow and QoS properties for each
component.

{\bf (2) Static analysis :} The interaction analysis tool checks the
coherence of the system architecture. For each partially compatible
interaction, a test to be performed at execution time is automatically
inserted into the CALICO {\tt debug} metamodel. For each incompatible
interaction, the architect is notified of the problem and he/she may
thus provide some modification of the application design. As long as
some incompatible interactions remain, the next steps of the
development process can not be reached. Once all of the problems are
fixed, the architect specifies the runtime platform on which the
application is to be executed and CALICO verifies that the
specifications do not go against the platform constraints in order to
make sure that the application can be indeed deployed on that specific
platform.

{\bf (3) Code generation :} If a component or service does not
already exist, then the {\tt generation} tool generates code skeleton such
that only business code needs to be provided by the developers.

{\bf (4) Instrumentalisation :} This step makes the link between the
static analysis and the dynamic checks of the application at runtime.
The {\tt instrumentation} tool takes the {\tt debug} model as input
and automatically instruments the application code to enable the
capture of the needed runtime information to complete the resolution
of the partially compatible interaction. This instrumentalisation
relies on an aspect-oriented approach and is independent of the
underlying platform.

{\bf (5) Instantiation :} The {\tt loader} instantiates the
application on the target platform as described by the architect's
structural model. Concretely, the running system is created
incrementally by calling the appropriate sequence of system
construction operations, such as creating/removing components and
connectors.

{\bf (6) Reification :} As the testers run the application in different
execution contexts, the instrumented application automatically reifies
any context changes and monitored information.

{\bf (7) Dynamic debugging :} During the debugging phase, the {\tt
  debug} tool analyzes the information reified by the running system
and triggers when needed the tests contained in the debugging model.
The architect is notified each time an error is detected, allowing
him/her to correct the application design. Other debugging action
rather than the notification action maybe chosen, such as logging the
information into a file, or executing a reconfiguration script that
will automatically modify the design and trigger the step 2 of the
process. This latter case may be useful to tune/test adaptation
policies for autonomic system.

{\bf (8) Evolution of the design :} The architect can modify the design with
respect to the debugging information if problems have occurred. They
can also adapt at anytime the design of the application to address new
user or application requirements. After any modification, the
development cycle iterates again starting at step 2.

\section{Illustrative Example}
\label{sec:example}

\begin{figure}[!ht]
	\centering
	\includegraphics[width=1.0\textwidth]{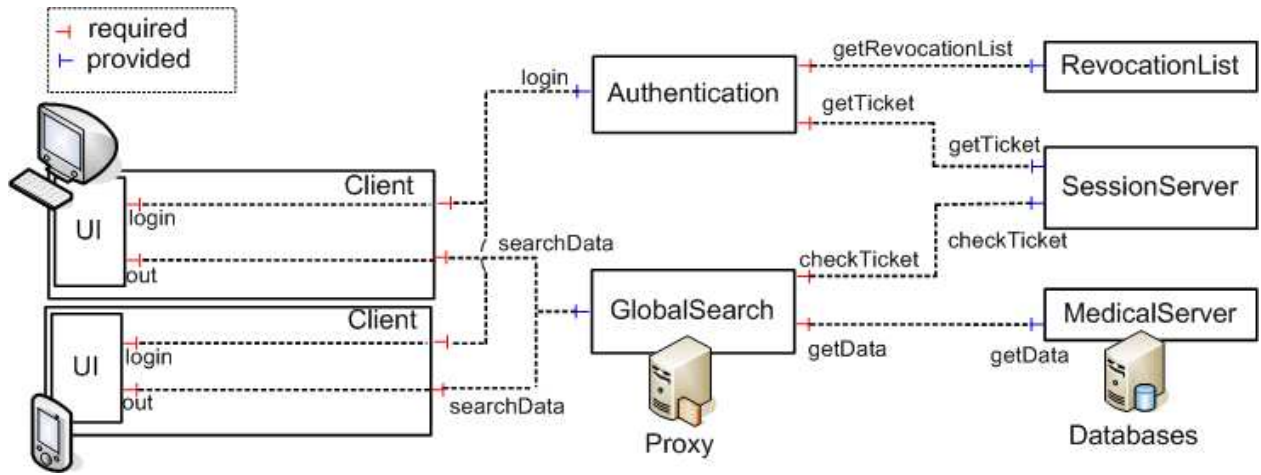}
	\caption{Structure of the PHR application}
	\label{fig:dmp}
\end{figure}

To illustrate the agile development process offered by CALICO, we use an
example of architecture in the context of the French Personal Health
Record system (PHR)~\cite{dmp}.  PHR is the French personal health
record system that is intended to provide health-care
professionals with the information needed for their patients care.
Figure~\ref{fig:dmp} represents a possible architecture of the PHR system.
All medical information, (such as biological analyses, X-rays,
medications, \textit{etc.}), will be stored in distributed {\tt databases}
 and will be made accessible through an on-line interface {\tt Client}.

In order to build a robust PHR application, architects need to be able
to express several application properties.  A first requirement of
this system architecture is related to authentication issues since not
everybody should have access to anybody's health records.  The
architecture of this system must thus provide some authentication
mechanism.  The {\tt Authentication} architecture element logs a
health-care professional in and returns a session ticket through the
functionality {\tt getTicket} that is offered by {\tt SessionServer}.
For security reason, the functionality {\tt getTicket} can be used
only by the element {\tt Authentication} to avoid that an
unauthenticated user get a session ticket.  Finally the session ticket
must be validated by the {\tt SessionServer} before retrieving any
medical data from the database.

Another requirement is the high reliability of the system.  Such
system has to be able to handle very heterogeneous medical
information, going from light-weight text records to gigabytes of
echographies.  Furthermore, the devices used to display this
information are also heterogeneous.  They range from desktop computers
with high-quality large-screen monitors and gigabyte network
connexions to simple PDAs with small screens and low-bandwidth GPRS
network connexion.  Handling such data in a reliable way is critical
because the system must be able to determine if a given data can be
displayed appropriately with no loss of information, as well as in
which time-frame, depending on the available resources and amount of
information to display. For example, a dataflow constraint may express
that medical data received on a terminal of type PDA Nokia N800 should
be less than 10 megaoctets. Another constraint can also specify that
only text or jpg documents can be read on that terminal.

Overall the constraints may evolve in time or just not reflect exactly
one given execution context. There is thus a need to iterate the whole
process to check if the declared application constraints can be all
checked, statically or dynamically.

{\bf (1)} The architect specifies the architecture and the properties
of the PHR application. For example, the first requirement mentioned
above can be specified using the structural contract metamodel.

{\bf (2)} The overall coherence of the constraints is statically
verified. A partially compatible interaction is detected between the
{\tt MedicalServer} and the {\tt PDS} since data sent by the server
could be greater than 10 megaoctets or in a different format than txt
or jpg. Accordingly CALICO adds some rules to validate in the {\tt
  debug} model. These rules specify that the size and the data type
must be captured at runtime.

{\bf (3)} Code skeletons are generated and developers can provide the
business code of each components.

{\bf (4)} Following the information contained in the {\tt debug}
model, the application is automatically modified to capture the size
and the type of the medical data that enters the PDA.

{\bf (5)} The application is deployed on the target application.

{\bf (6)} At this step different execution contexts are tested. One
may consists in the use of the PHR application by a druggist, who
typically uses the PHR only to consult text documents. Another test
scenario considers a radiologist. During the test scenario execution,
monitored informations are reified.

{\bf (7)} The {\tt debug} tool resumes the interaction compatibility
checks that were partially compatible. In the case of the druggist, no
error is detected, whereas for the radiologist, the analyse indicates
that the data are too large for the PDA.

{\bf(8)} The architect can accordingly modify the application design
by inserting a new component {\tt DataConverter} between the PDA and
the {\tt GlobalSearch} component in order to reduce the size of a too
large radiography.

The whole process is then iterated again. If no error is detected
statically, the new component {\tt DataConverter} is automatically
integrated into the already deployed application, and new test
scenarios may be executed.

\section{Conclusion and Current Implementation Status}
\label{sec:conclusion}

CALICO is a model-based framework that enables the design and debug of
systems in an iterative and incremental way, bridging a little more
the gap between component-based software engineering and agile
development approaches. Our framework is generic and highly
extensible. All metamodels for specifying the structure, the
application properties and the adaptation rules are independent of any
underlying platform. This enables architects to perform various
architecture analyses on their applications even if the underlying
component or service framework does not provide any verification
tools.

The current implementation of CALICO is developed in Java. All CALICO
metamodels are implemented with the Eclipse Modeling Framework (EMF).
A graphical editor, implemented with the Graphical Modeling Framework
(GMF), enables the architect to edit the model during the whole
development cycle. 

We have integrated several existing tools to verify the coherence of
the component interactions in term of structural, behavioral, dataflow
and quality of service properties. Structural constraints are
expressed in OCL~\cite{ocl}, using the EMF-OCL library. Behavioral
specifications are based on existing process algebra, such as
CSP~\cite{csp}, FSP~\cite{fsp}, SFSP. The current implementation uses
the Fractal behavioral protocol checker~\cite{fractalbpc} to verify
that a given component composition does not introduce a deadlock. We
have developed a dataflow analysis based on the algorithm of constant
propagation in partial program validation~\cite{Gary}. The QoS
metamodel has been inspired by the QML~\cite{qml} and WSLA~\cite{wsla}
approaches. The associated analysis is based on prediction of quality
property in a workflow of WEB services~\cite{10.1109/ICEBE.2007.18}.
Furthermore, application instrumentation has been implemented with
Spoon~\cite{spoon}. The sensor framework Wildcat~\cite{wildcat} has
been integrated in CALICO. Our current implementation supports four
component platforms (Fractal~\cite{FractalJava},
OpenCCM~\cite{opencom}, OpenCOM~\cite{BRICLET:2004:HAL-00003287:1} and
FraSCAti~\cite{frascati}) and one service-oriented platform(Web
services~\cite{soa}). CALICO has been carefully designed to allow new
extensions in terms of support for new platforms, new QoS sensors and
new kinds of debugging actions.

We have performed benchmarks on our implementation and showed that CALICO is usable to
design reliable large systems up-to 10000 components, which is the maximum
load of most runtime platforms.

CALICO is still being developed, to support more extensions. The
current implementation is freely available at
http://calico.gforge.inria.fr.

\end{document}